\documentclass[aps,prd,twocolumn,showpacs,floatfix]{revtex4}

\usepackage{amsmath}
\usepackage{amssymb}
\usepackage{graphicx}
\usepackage{bm}

\begin{document}
\title{Wormhole geometries with conformal motions}

\author{Christian G.~B\"ohmer}
\email{c.boehmer@ucl.ac.uk} \affiliation{Department of
Mathematics, University College London,
             Gower Street, London, WC1E 6BT, UK}

\author{Tiberiu Harko}
\email{harko@hkucc.hku.hk} \affiliation{Department of Physics and
Center for Theoretical
             and Computational Physics, The University of Hong Kong,
             Pok Fu Lam Road, Hong Kong}

\author{Francisco S.~N.~Lobo}
\email{francisco.lobo@port.ac.uk}
\affiliation{Institute of Cosmology \& Gravitation,
             University of Portsmouth, Portsmouth PO1 2EG, UK}
\affiliation{Centro de Astronomia e Astrof\'{\i}sica da
             Universidade de Lisboa, Campo Grande, Ed. C8 1749-016 Lisboa,
             Portugal}

\date{\today}

\begin{abstract}

Exact solutions of traversable wormholes were recently found under
the assumption of spherical symmetry and the existence of a {\it
non-static} conformal symmetry. In this paper, we verify that in
the case of the conformally symmetric spacetimes with a non-static
vector field generating the symmetry, the conformal factor $\psi $
can be physically interpreted in terms of a measurable quantity,
namely, the tangential velocity of a massive test particle moving
in a stable circular orbit in the spacetime. Physical properties
of the rotational velocity of test particles and of the redshift
of radiation emitted by ultra-relativistic particles rotating
around these hypothetical general relativistic objects are further
discussed. Finally, specific characteristics and properties of
gravitational bremsstrahlung emitted by charged particles in
geodesic motion in conformally symmetric wormhole geometries are
also explored.

\end{abstract}


\maketitle

\section{Introduction}

Wormholes are hypothetical shortcuts in spacetime
\cite{Morris:1988cz,Visser}, and are primarily useful as
``gedanken-experiments'' and as a theoretician's probe of the
foundations of general relativity. These solutions are a specific
example in solving the Einstein field equation in the reverse
direction, namely, one first considers an interesting spacetime
metric, then finds the matter source responsible for the
respective geometry. In this manner, it was found that some of
these solutions possess a peculiar property, namely ``exotic
matter,'' involving a stress-energy tensor that violates the null
energy condition, and a number of specific solutions have been
found (see Ref. \cite{solutions} and references therein, and Ref.
\cite{Lobo:2007zb} for a recent review). These geometries also
allow closed timelike curves, with the respective causality
violations \cite{Morris:1988tu}.

A more systematic approach in searching for exact solutions,
namely, by assuming spherical symmetry and the existence of a
\textit{non-static} conformal symmetry, was recently considered in
\cite{Boehmer:2007rm}. Suppose that the vector $\mathbf{\xi }$
generates the conformal symmetry, then the metric $\mathbf{g}$ is
conformally mapped onto itself along $\mathbf{\xi }$. This is
translated by the following relationship
\begin{equation}
\mathcal{L}_{\mathbf{\xi }}\,\mathbf{g}=\psi \,\mathbf{g}\,,
\label{Lieconf}
\end{equation}
where $\mathcal{L}$ is the Lie derivative operator and $\psi $ is
the conformal factor. Note that neither $\xi $ nor $\psi $ need to
be static even though one considers a static metric. A wide
variety of solutions with the exotic matter restricted to the
throat neighborhood and with a cut-off of the stress-energy tensor
at a junction interface were deduced, and particular
asymptotically flat geometries were also found. The specific
solutions were deduced by considering choices for the form
function, an equation of state relating the energy density and the
anisotropy, and phantom wormhole geometries were also explored.

The assumption of a static conformal symmetry, i.e., with a static
vector $\xi$, considered in \cite{Herrera,Maartens:1989ay} was
found responsible for the singular solutions at the center.
However, we emphasize that this is not problematic to wormhole
physics, due to the absence of a center \cite{Boehmer:2007rm}.
Indeed, the case of a non-static conformal symmetry $\xi$
generating vector field $\xi$ does not yield a singularity at
$r=0$ and a wide variety of solutions were found
\cite{Maartens:1989ay}. Therefore, since the origin is an allowed
point, the analysis is in fact more general than wormholes. In
this context, it is interesting to note that the physical
properties and characteristics of conformal symmetries have also
been applied to a wide variety of geometries in the literature
\cite{Mak:2003kw,Harko:2004ui,Mak:2004hv}, in particular, an exact
analytical solution describing the interior of a charged strange
quark star was found~ \cite{Mak:2003kw}; solutions were also
explored in braneworlds~\cite{Harko:2004ui}; and also in the
context of the galactic rotation curves~\cite{Mak:2004hv}.

It is the purpose of this paper to consider the behavior of some
observationally relevant physical quantities in the conformally
symmetric wormhole geometry analyzed in detail in
\cite{Boehmer:2007rm}. As a first step we consider the behavior of
the massive test particles in stable circular orbits around the
wormhole. Under the assumption of spherical symmetry we derive the
basic equation describing the tangential velocity of the particle,
which can be obtained as a function of the derivative with respect
to the radial coordinate of the redshift function only. On the
other hand, because the derivative of the redshift function can be
expressed in terms of the conformal factor describing the basic
{\it geometrical} properties of the wormhole, it follows that the
conformal factor can be obtained directly from the tangential
velocity. The conformal factor also determines the form function.
The tangential velocity of particles in circular orbits around
wormholes could be determined, at least in principle, from
astronomical observations of the frequency shifts (both blue and
red) of the light emitted by the charged particles orbiting the
wormhole. Therefore this opens the possibility of a direct
observational determination of the geometry of the conformally
symmetric wormholes.

Another observationally important physical parameter is the total
power radiated by charged particles, either spiraling towards the
wormhole in circular orbits, or falling on radial paths. The
process of electromagnetic radiation emission due to the
gravitational acceleration of charged particles in geodesic motion
is called gravitational bremsstrahlung. The measurement of this
power from astronomical/astrophysical observations would allow the
determination of the conformal factor, thus allowing a direct test
of the wormhole geometry.

The present paper is organized as follows. We review the
properties of the conformally symmetric wormholes in Section
\ref{secII}. The relation between the tangential velocity of
particles in stable circular orbits and the geometric properties
of the spacetime is considered in Section \ref{secIII}. We analyze
the behavior of the rotational velocity of test particles and of
the frequency shifts of light in Section \ref{secIV}. In Section
\ref{secV}, we explore some physical properties and
characteristics of gravitational bremsstrahlung emitted by charged
particles in geodesic motion in conformally symmetric wormhole
geometries. Finally, we discuss and conclude our results in
Section \ref{secVI}.

\section{Conformal symmetry and  wormhole geometry}\label{secII}

An alternative approach in searching for exact wormhole solutions
\cite {Boehmer:2007rm} is based on the assumption that the
spherically symmetric static spacetime  possesses a conformal
symmetry \cite {Herrera,Maartens:1989ay}. It is interesting to
note, as mentioned in the Introduction, that neither $\xi $ nor
$\psi $ need to be static even though one considers a static
metric, and therefore Eq. (\ref{Lieconf}) takes the following form
\begin{equation}
g_{\mu \nu ,\alpha }\,\xi ^{\alpha }+g_{\alpha \nu }\,\xi ^{\alpha
}{}_{,\mu }+g_{\mu \alpha }\,\xi ^{\alpha }{}_{,\nu }=\psi
\,g_{\mu \nu }\,. \label{Lie}
\end{equation}
As in Ref. \cite{Boehmer:2007rm}, we follow closely the
assumptions made in Ref. \cite{Maartens:1989ay}, where the
condition
\begin{equation}
\mathbf{\xi }=\alpha (t,r)\,\partial _{t}+\beta (t,r)\,\partial
_{r}\,, \label{Lie3}
\end{equation}
is considered, and the conformal factor is static, i.e., $\psi
=\psi (r)$.

The spacetime metric that will be considered, representing a
spherically symmetric and static wormhole, is given by
\begin{equation}
ds^2=-e ^{2\Phi(r)}\,dt^2+\frac{dr^2}{1- b(r)/r}+ r^2 d\Omega^2
\,, \label{metricwormhole}
\end{equation}
where $d\Omega ^{2}=d\theta ^{2}+\sin ^{2}\theta d\phi ^{2}$, and
$\Phi(r)$ and $b(r)$ are arbitrary functions of the radial
coordinate, $r$, denoted as the redshift function, and the form
function, respectively \cite {Morris:1988cz}. The radial
coordinate $r$ possesses a minimum value at $r_0$, representing
the location of the throat of the wormhole, where $b(r_0)=r_0$,
and consequently has the following range $r\in [r_0,+\infty)$. The
proper circumference of a circle of fixed $r$ is given by $2\pi
r$. The redshift function $\Phi(r)$ is finite throughout the range
of interest in order to avoid the absence of event horizons. The
form function is also constrained by the flaring-out condition,
$(b^{\prime}r-b)/b^2<0$, which reduces to $b^{\prime}(r_0)<1$ at
the throat.

Taking into account metric (\ref{metricwormhole}), then Eq.
(\ref{Lie}), without a loss of generality (see Ref.
\cite{Boehmer:2007rm} for details), provides the following
solutions
\begin{equation*}
\xi =\frac{1}{2}kt\,\partial _{t}+\frac{1}{2}\psi (r)r\,\partial
_{r}\,
\end{equation*}
and
\begin{eqnarray}
b(r) &=&r[1-\psi ^{2}(r)]\,,  \label{Liesolution2b} \\
\Phi (r) &=&\frac{1}{2}\ln (C^{2}r^{2})-k\int \frac{dr^{\prime
}}{r^{\prime }\psi (r^{\prime })}\,,  \label{Liesolution2c}
\end{eqnarray}
respectively, where $k$ $C$ are constants. An interesting feature
of these solutions that immediately stands out, by taking into
account Eq. (\ref
{Liesolution2b}), is that the conformal factor is zero at the throat, i.e., $%
\psi (r_{0})=0$.

Note that the solutions given by Eqs. (\ref{Liesolution2b}) and
(\ref{Liesolution2c}) impose the following condition, relating the
form and redshift functions
\begin{equation}
\Phi'(r)=\frac{1}{r}\left(1-\frac{k}{\sqrt{1-b(r)/r}}\right) \,.
    \label{relationPhib}
\end{equation}
Note that this relationship places a strong constraint on the
specific choices of the wormhole geometries. A wide range of
specific solutions were deduced in Ref. \cite{Boehmer:2007rm}, by
considering choices for the form function. Note that, in
principle, one may also impose interesting choices for the
redshift function, and consequently deduce the form function and
the conformal factor.

It is rather important to emphasize that specific examples of
asymptotically flat spacetimes were found in
\cite{Boehmer:2007rm}, by considering the case of $k=1$ and
normalizing the integration constant in Eq. (\ref{Liesolution2c})
by imposing the value $C^2=2$. Thus, in the analysis that follows,
one may consider asymptotically flat spacetimes simply by imposing
the above values for the respective constants.

The existence of conformal motions imposes strong constraints on
the wormhole geometry, so that the stress-energy tensor components
are written solely in terms of the conformal function
\cite{Boehmer:2007rm}, and take the following form
\begin{eqnarray}
\rho(r)&=&\frac{1}{\kappa^2r^2}\left(1-\psi^2-2r\psi\psi^{\prime}\right)
\,,
\label{rhoWH2} \\
p_r(r)&=&\frac{1}{\kappa^2r^2} \left(3\psi^2-2k\psi-1\right) \,,
\label{prWH2} \\
p_t(r)&=&\frac{1}{\kappa^2r^2}\left(\psi^2-2k\psi+k^2+2r\psi\psi^{\prime}%
\right) \,,  \label{ptWH2}
\end{eqnarray}
where $\kappa^2=8\pi$; $\rho(r)$ is the energy density, $p_r(r)$
is the radial pressure, and $p_t(r)$ is the lateral pressure
measured in the orthogonal direction to the radial direction. Note
that the conservation of the stress-energy tensor provides the
following relationship
\begin{equation}
p_r^{\prime}=\frac{2}{r}\,(p_t-p_r)-(\rho +p_r)\,\Phi ^{\prime}\,.
\label{prderivative}
\end{equation}

The NEC violation, for this case, is given by
\begin{equation}
\rho (r)+p_{r}(r)=\frac{1}{\kappa ^{2}r^{2}}\,\left[ 2\psi (\psi
-k)-r(\psi ^{2})^{\prime }\right] \,,  \label{NECviol2}
\end{equation}
which evaluated at the throat imposes the following condition
$(\psi ^{2})^{\prime }>0$.

\section{Stable circular orbits in static and spherically symmetric spacetimes}
\label{secIII}

In the case of the conformally symmetric spacetimes with a
non-static vector field generating the symmetry, the conformal
factor $\psi $ can be physically interpreted in terms of a
measurable quantity, the tangential velocity of a massive test
particle moving in a stable circular orbit in the spacetime
described by the line element given by Eq.~(\ref{metricwormhole}).

To verify this, consider the Lagrangian $\mathcal{L}$ for a
massive test particle, which reads
\begin{equation}
\mathcal{L}=\frac{1}{2}\left[ -e^{2\Phi }\dot{t}^{2}+\frac{\dot{r}^{2}}{1-b/r%
}+r^{2}\dot{\Omega}^{2}\right] ,  \label{lag}
\end{equation}
where the overdot denotes differentiation with respect to the
affine parameter $s$. Since the metric tensor coefficients do not
explicitly depend on $t$ and $\Omega $, the Lagrangian~(\ref{lag})
yields the following conserved quantities (generalized momenta):
\begin{equation}
-e^{2\Phi (r)}\dot{t}=E,\qquad r^{2}\dot{\Omega}=L,  \label{cons}
\end{equation}
where $E$ is related to the total energy of the particle and $L$
to the total angular momentum. With the use of the conserved
quantities, we obtain from Eq.~(\ref{lag}) the geodesic equation
for massive particles in the form
\begin{equation}
e^{2\Phi }\left( 1-\frac{b}{r}\right) ^{-1}\dot{r}^{2}+e^{2\Phi }\left( 1+%
\frac{L^{2}}{r^{2}}\right) =E^{2}\,.  \label{geod1}
\end{equation}

This equation shows that the radial motion of the particles on a
geodesic is
the same as that of a particle with position dependent mass and with energy $%
E^{2}/2$ in ordinary Newtonian mechanics moving in the effective
potential
\begin{equation}
V_{eff}\left( r\right) =e^{2\Phi }\left(
\frac{L^{2}}{r^{2}}+1\right) .
\end{equation}

For the case of the motion of particles in circular and stable
orbits the effective potential must satisfy the following
conditions: a) $\dot{r}=0$, representing circular motion; b)
$\partial V_{eff}/\partial r$ $=0$,
providing extreme motion; c) $\partial ^{2}V_{eff}/\partial r$ $%
^{2}|_{extr}>0$, translating a stable orbit. Conditions a) and b)
immediately provide the conserved quantities as
\begin{equation}
E^{2}=e^{2\Phi }\left( 1+\frac{L^{2}}{r^{2}}\right) ,
\label{cons1}
\end{equation}
and
\begin{equation}
\frac{L^{2}}{r^{2}}=r\Phi ^{\prime }e^{-2\Phi }E^{2},
\label{cons2}
\end{equation}
respectively. Equivalently, these two equations can be rewritten
as
\begin{equation}
E^{2}=\frac{e^{2\Phi }}{1-r\Phi ^{\prime }}\,,\qquad
L^{2}=\frac{r^{3}\Phi ^{\prime }}{1-r\Phi ^{\prime }}\,.
\end{equation}

The line element, given by Eq. (\ref{metricwormhole}), can be
rewritten in terms of the spatial components of the velocity
\cite{LaLi}, where
\begin{equation}
v^{2}=e^{-2\Phi }\left[ \frac{1}{1-b(r)/r}\left(
\frac{dr}{dt}\right) ^{2}+r^{2}\left( \frac{d\Omega }{dt}\right)
^{2}\right] .
\end{equation}

For a stable circular orbit $\dot{r}=0$, the tangential velocity
of the test particle can be expressed as
\begin{equation}
v_{\mathrm{tg}}^{2}=r^{2}e^{-2\Phi }\left( \frac{d\Omega
}{dt}\right) ^{2}=e^{-2\Phi
}r^{2}\dot{\Omega}^{2}/\dot{t}^{2}=e^{2\Phi
}\frac{L^{2}}{r^{2}E^{2}}\,. \label{vtgbr}
\end{equation}
By using the constants of motion, we obtain the expression of the
tangential velocity of a test particle in a stable circular orbit,
given by
\begin{equation}
v_{\mathrm{tg}}^{2}=r\Phi ^{\prime }.  \label{vtg}
\end{equation}

Thus, the rotational velocity of the test body is determined by
the redshift function only. This expression which relates one of
the metric components to the tangential velocity is an exact
general relativistic expression valid for static and spherically
symmetric spacetimes. Note that for generic redshift functions,
where $\Phi ^{\prime }<0$, there are no stable circular orbits.
The expression also imposes that $0\leq r\Phi ^{\prime }<1$. The
specific case of a constant redshift function imposes that
$v_{\mathrm{tg}}^{2}=0$ for all values of $r$.

As a second consistency condition we require that the timelike
circular geodesics in the wormhole geometry be stable. Let
$r_{eq}$ be a circular orbit and consider a perturbation of it of
the form $r=r_{eq}+\delta $, where $\delta \ll r_{eq}$
\cite{La03}. Taking expansions of $V_{eff}\left( r\right) $ and
$1/\left( 1-b(r)/r\right) $ about $r=r_{eq}$, it follows from Eq.
(\ref{geod1}) that
\begin{equation}
\ddot{\delta}+\frac{1}{2}e^{-2\Phi \left( r_{eq}\right) }\left[ 1-\frac{%
b\left( r_{eq}\right) }{r_{eq}}\right] V_{eff}^{\prime \prime
}\left( r_{eq}\right) \delta =0.
\end{equation}

The condition for stability of the simple circular orbits requires $%
V_{eff}^{\prime \prime }\left( r_{eq}\right) >0$ \cite{La03}. This
gives
\begin{equation}
\left. \left( 3\Phi ^{\prime }+r\Phi ^{\prime \prime }-2r\Phi
^{\prime 2}\right) \right| _{r=r_{eq}}>0.
\end{equation}

In terms of the tangential velocity the stability condition can be
reformulated as
\begin{equation}\label{equil}
\left.\left[\frac{d}{dr}v_{\mathrm{tg}}^{2}+\frac{2v_{\mathrm{tg}}^{2}
\left(1-{\mathrm{tg}}^{2}\right)}{r}\right]\right|
_{r=r_{eq}}>0.
\end{equation}

\section{Tangential velocity and redshift in conformally symmetric wormhole
geometry}\label{secIV}

In the case of the motion of a test particle in a conformally
symmetric, static spherically symmetric space-time, with a
non-static vector field generating the symmetry, the metric
coefficient $\exp \left( 2\Phi \right) $ is given by Eq.
(\ref{Liesolution2c}). Therefore for the angular velocity we find
the simple expression
\begin{equation}
v_{\mathrm{tg}}^{2}=1-\frac{k}{\psi }.  \label{tg}
\end{equation}
Equation (\ref{tg}) gives a direct physical interpretation of the
conformal
factor $\psi $ in terms of the tangential velocity, $\psi =k/\left( 1-v_{%
\mathrm{tg}}^{2}\right) $. From Eq.~(\ref{tg}) it follows that the
general, physically acceptable, range of the parameter $\psi $ is
$\psi \in \lbrack k,\infty )$, corresponding to a variation of the
tangential velocity between zero and the speed of light. Note that
this is equivalent to stating that the expression is only valid
for stable circular orbits, so that the condition $0<k/\psi <1$ is
imposed. This condition, in particular, excludes the throat,
$r_{0}$ (recall that $\psi (r_{0})=0)$, which reflects the absence
of stable circular orbits around the throat. Note that as the
relationship (\ref{relationPhib}) places a strong constraint on
the specific choices of the wormhole geometries, and as the
emphasis is on physically measurable quantities the conformal
symmetry places severe restrictions on the range of stable orbits.

This interesting physical phenomenon can be explained on the
grounds that in the wormhole geometry the instability develops due
to the violation of the null energy condition $\rho+p_r<0$. This
reflects that gravity becomes repulsive in these regimes, and the
repulsive force destabilizes the orbit, where the angular momentum
cannot balance with the gravitational interaction.

In terms of the conformal factor the stability condition given by
Eq. (\ref{equil}) can be reformulated as
\begin{equation}
\left.k\left[\frac{d\psi }{dr}+\frac{2}{r}\left(\psi
-k\right)\right]\right|_{r=r_{eq}}>0.
\end{equation}

On the other hand, the form function $b(r)$ can also be expressed
as a function of the tangential velocity only:
\begin{equation}
b(r)=r\left[ 1-\frac{k^{2}}{\left( 1-v_{\mathrm{tg}}^{2}\right)
^{2}}\right] .  \label{form2}
\end{equation}

The condition of the NEC violation can be formulated in terms of
the tangential velocity of test particles as
\begin{equation}
\rho \left( r\right) +p_{r}\left( r\right) =\frac{1}{r^{2}}\left[ \frac{2v_{%
\mathrm{tg}}^{2}}{1-v_{\mathrm{tg}}^{2}}-r\frac{d}{dr}\frac{1}{\left( 1-v_{%
\mathrm{tg}}^{2}\right) ^{2}}\right] <0. \label{NEC2}
\end{equation}
We emphasize that Eqs. (\ref{form2}) and (\ref{NEC2}) are only
valid  for stable circular orbits in the range $0<k/\psi <1$.
Thus, they do not apply, in particular, to the wormhole throat
$r_0$, where the conformal factor obeys $\psi(r_0)=0$.

The relationship (\ref{NEC2}) is equivalent with the following
condition which must be satisfied by the tangential velocity of a
test particle in a wormhole geometry:
\begin{equation}
\frac{dv_{\mathrm{tg}}}{dr}>\frac{1}{2r}v_{\mathrm{tg}}\left(
1-v_{\mathrm{tg}}^{2}\right) ^{2}.
\end{equation}
Note that this relationship is only valid if we impose that exotic
matter is threaded throughout the wormhole. By integrating this
equation we obtain the physical condition defining the conformally
symmetric wormhole geometry as
\begin{equation}
v_{\mathrm{tg}}^{2}>\frac{r/R_{0}}{1+r/R_{0}},
\end{equation}
where $R_{0}>0$ is an arbitrary length scale (a constant of
integration).

Finally, we consider the possibility of observationally testing
wormhole geometries via the observation of the frequency shifts of
the light emitted by particles orbiting wormholes. Depending on
the direction of motion with respect to an observer located at
infinity, the radiation emitted by particles moving in circular
orbits on both sides of the central region of the wormhole will be
successively red shifted (in the case of the recessional motion)
and blue shifted (in the case of the particle approaching the
observer), respectively. Consider two observers $O_{E}$ and
$O_{\infty }$, with four-velocities $u_{E}^{\mu }$ and $u_{\infty
}^{\mu }$, respectively. Observer $O_{E}$ corresponds to the light
emitter (i.e., to the particles in stable circular orbits orbiting
around a wormhole and located at a point $P_{E}$ of the
space-time), and $O_{\infty }$ represents the detector at point
$P_{\infty }$, located far from the emitter, and that can be
idealized to correspond to ``spatial infinity''.

The light signal travels to the observer on null geodesics with
tangent vector $k^{\mu }$. We may restrict $k^{\mu }$ to lie on
the equatorial plane $\theta =\pi /2$ of the wormhole, without a
significant loss of generality, and evaluate the frequency shift
for a light signal emitted from $O_{E}$ in a circular orbit and
detected by $O_{\infty }$. The frequency shift associated to the
emission and detection of the light signal is given by
\begin{equation}
z=1-\frac{\omega _{E}}{\omega _{\infty }},
\end{equation}
where $\omega _{I}=-k_{\mu }u_{I}^{\mu }$, and the index $I$
refers to emission ($I=E$) or detection ($I=\infty $) at the
corresponding space-time point \citep{La03}. Two frequency shifts,
corresponding to maximum and minimum values, are associated with
light propagation in the same and opposite direction of motion of
the emitter, respectively. Such shifts are frequency shifts of a
receding or approaching particle, respectively. Using
the constancy along the geodesic of the product of the Killing field $%
\partial /\partial t$ with a geodesic tangent gives the expressions of the
two shifts as \cite{La03}
\begin{equation}
z_{\pm }=1-e^{\Phi _{\infty }-\Phi \left( r\right) }\frac{1\mp
\sqrt{r\Phi ^{\prime }}}{\sqrt{1-r\Phi ^{\prime }}}=1-e^{\Phi
_{\infty }-\Phi \left( r\right) }\frac{1\mp
v_{\mathrm{tg}}}{\sqrt{1-v_{\mathrm{tg}}^{2}}},
\end{equation}
respectively, where $\exp \left[ 2\Phi \left( r\right) \right] $
represents
the value of the metric potential at the radius of emission $r$ and $\exp %
\left[ 2\Phi _{\infty }\right] $ represents the corresponding value of $\exp %
\left[ 2\Phi \left( r\right) \right] $ for $r\rightarrow \infty $.

In terms of the conformal factor $\psi $ the two shifts can be
written as
\begin{equation}
z_{\pm }=1-e^{\Phi _{\infty }-\Phi \left( r\right) }\sqrt{\frac{\psi }{k}}%
\left( 1\mp \sqrt{1-\frac{k}{\psi }}\right) .
\end{equation}

It is convenient to define two other quantities $z_{D}=\left(
z_{+}-z_{-}\right) /2$ and $z_{A}=\left( z_{+}+z_{-}\right) /2$,
given by
\begin{eqnarray}
z_{D}\left( r\right) =e^{\Phi _{\infty }-\Phi \left( r\right) }\frac{\sqrt{%
r\Phi ^{\prime }}}{\sqrt{1-r\Phi ^{\prime }}}=\nonumber\\
 e^{\Phi
_{\infty }-\Phi \left( r\right)
}\frac{v_{\mathrm{tg}}}{\sqrt{1-v_{\mathrm{tg}}^{2}}}= e^{\Phi
_{\infty }-\Phi \left( r\right) }\sqrt{\frac{\psi }{k}-1},
\end{eqnarray}
\begin{equation}
z_{A}\left( r\right) =1-\frac{e^{\Phi _{\infty }-\Phi \left( r\right) }}{%
\sqrt{1-r\Phi ^{\prime }}}=1-\sqrt{\frac{\psi }{k}}e^{\Phi
_{\infty }-\Phi \left( r\right) },
\end{equation}
respectively, which can be easily connected to the astronomical observations %
\cite{La03}. $z_{A}$ and $z_{D}$ satisfy the relation $\left(
z_{A}-1\right) ^{2}-z_{D}^{2}=\exp \left[ \Phi _{D}-\Phi \left( r\right) %
\right] $, and thus in principle both $\exp \left[ 2\Phi \left( r\right) %
\right] $ and $\psi \left( r\right) $ could be obtained directly
from the observations. Finally, it should be noted that the
tangential velocity can be expressed solely in terms of the shifts
\begin{equation}
v_{\mathrm{tg}}^{2} = \frac{z_D}{1-z_A}.
\end{equation}

\section{Gravitational bremsstrahlung from charged
particles}\label{secV}

Charged particles moving on a geodesic path in a gravitational
field {\it do emit} an electromagnetic, bremsstrahlung type
radiation, as expected from classical electrodynamics, where
acceleration is the source of the electromagnetic radiation. This
phenomenon was pointed out first in \cite{WiBr60}, and further
investigated and corrected in \cite{Hobbs}, where the process was
called electro-gravitic bremsstrahlung. However, an alternative
name for this specific type of electromagnetic radiation, which is
in a much wider use presently, is {\it gravitational
bremsstrahlung}. The computation of the high-frequency radiation
emitted by freely falling particles moving in circular geodesic
orbits in a spherically symmetric gravitational field, and in the
Kerr geometry, respectively, was performed in \cite{Mis73}. The
properties of gravitational bremsstrahlung radiation were
considered in detail, in different physical situations and
geometrical frameworks, in \cite{rad}.

In the present Section, we calculate the total power emitted by
charged particles in geodesic motion in the conformally symmetric
wormhole geometry. The four-momentum $dp^{\mu }$ radiated by a
charged particle with a four-velocity $u^{\mu }=\left( \gamma
/\sqrt{-g_{00}},\gamma v^{i }\right) $, where $\gamma
=1/\sqrt{1-v^{2}}$ and $v^{2}=\gamma _{ij}v^{i}v^{j}$, with
$\gamma _{ij}=g_{ij}$ the spatial three-dimensional metric tensor,
is given by \cite{LaLi}
\begin{equation}
dp^{\mu }=\frac{2}{3}e^{2}\frac{du^{\alpha }}{ds}\frac{du_{\alpha
}}{ds}dx^{\mu }.
\end{equation}

The motion of the particle in the wormhole geometry takes places
along the geodesics of the space-time, given by
\begin{equation}
\frac{du^{\alpha }}{ds}+\Gamma _{\rho \sigma }^{\alpha }u^{\rho
}u^{\sigma }=0,
\end{equation}
and
\begin{equation}
\frac{du_{\alpha }}{ds}=\frac{1}{2}\frac{\partial g_{\gamma \delta }}{%
\partial x^{\alpha }}u^{\gamma }u^{\delta },
\end{equation}
respectively \cite{LaLi}. Therefore, the total four-momentum
radiated by a charged particle in geodesic motion around a
wormhole is given by
\begin{equation}
\Delta p^{\mu }=-\frac{1}{3}e^{2}\int \Gamma _{\rho \sigma
}^{\alpha }\frac{\partial g_{\gamma \delta }}{\partial x^{\alpha
}}u^{\rho }u^{\sigma } u^{\gamma }u^{\delta } dx^{\mu }.
\end{equation}

In particular, for a static metric we obtain the total radiated
energy $\Delta E $ as
\begin{equation}
\Delta E=-\frac{1}{3}e^{2}\Gamma _{\rho \sigma }^{\alpha
}\frac{\partial g_{\gamma \delta }}{\partial x^{\alpha }}u^{\rho
}u^{\sigma } u^{\gamma }u^{\delta } \Delta t,
\end{equation}
and hence the total radiated power $P=\Delta E/\Delta t$ is given
by
\begin{equation}
P=-\frac{1}{3}e^{2}\Gamma _{\rho \sigma }^{\alpha }\frac{\partial
g_{\gamma \delta }}{\partial x^{\alpha }}u^{\rho }u^{\sigma
}u^{\gamma }u^{\delta }.
\end{equation}

Consider first the case of the geodesic motion of a charged
particle in the polar plane, $\phi =0$, without a significant loss
of generality, of the wormhole. The only component of the
three-dimensional velocity is $v^{\theta }=\exp(-\Phi )d\theta
/dt$, and the square of the velocity is $v^2=r^2\exp\left(-2\Phi
\right)\left(d\theta /dt\right)^2=v_{\mathrm{tg}}^2$. Hence, we
have $v^{\theta }=v_{\mathrm{tg}}/r$.

Using $u^{\mu}=\left( \gamma e^{-\Phi},0,\gamma
v_{\mathrm{tg}}/r,0\right) $, we obtain the radiated power by a
charged particle in geodesic rotation in the conformally symmetric
wormhole geometry as
\begin{equation}
P=\frac{2e^2\gamma^4}{3}\left(1-\frac{b}{r} \right)\left(\Phi^{\prime}-%
\frac{v_{\mathrm{tg}}^2}{r} \right)^2.
\end{equation}

Once the particle radiates, its energy is not conserved, and
therefore Eq. (\ref{vtg}) cannot be used, as for stable circular
orbits we have $v_{\mathrm{tg}}^2=r\Phi'$, and therefore $P=0$.
The tangential velocity is an arbitrary input parameter. Moreover,
for large distances from the wormhole, we can neglect the term
$v_{\mathrm{tg}}^2$, thus obtaining for the radiated power
\begin{equation}\label{radap}
P\approx \frac{2e^{2}\gamma ^{4}}{3}\left( 1-\frac{b}{r}\right)
\Phi ^{\prime 2}=\frac{2e^{2}\gamma ^{4}}{3r^2}\left( \psi
-k\right) ^{2},
\end{equation}
where we have also expressed the conformal factor $\psi (r)$ in
terms of  form function $b(r)$ by using the relation $\psi
(r)=\sqrt{1-b(r)/r}$.

In the case of a purely radial geodesic motion of a charged
massive particle in the plane $\phi =0$ the only non-zero
component of the three velocity is $v^r=\exp\left(-\Phi
\right)dr/dt$, with the square given by
\begin{equation}
v^2=e^{-2\Phi
}\left(1-\frac{b(r)}{r}\right)^{-1}\left(\frac{dr}{dt}\right)^2.
  \label{velocity}
\end{equation}
Therefore we obtain
\begin{equation}
v^r=\sqrt{1-\frac{b(r)}{r}}v=\psi (r)v,
\end{equation}
and consequently $u^{\mu}=\left( \gamma e^{-\Phi},\gamma
\sqrt{1-b(r)/r}v,0,0\right) $.

Note that Eq. (\ref{velocity}) may also be expressed in terms of
the constant of motion $E$, given by
\begin{equation}
v^2=1-\frac{e^{2\Phi}}{E^2}.
\end{equation}

Finally, we obtain for the radiated power of a charged particle
radially falling into the wormhole the expression
\begin{equation}
P=\frac{2e^{2}\gamma ^{4}}{3}\left( 1-\frac{b}{r}\right) \left[
\Phi ^{\prime 2}-\frac{\left(b-rb^{\prime}\right)^2 }{4r^{4}\left(
1-b/r\right) ^{2}}v^{4}\right] .
\end{equation}

In terms of the conformal factor the power emitted for a particle
in radial motion is
\begin{equation}
P=\frac{2e^{2}\gamma ^{4}}{3r^{2}}\left[ \left( \psi-k \right)
^{2}-\left(r \psi ^{\prime}v^{2} \right)^2\right] .
\end{equation}

In the limit of large $r$ we can neglect the term proportional to
$v^2$ in the expression of the power, and we obtain again Eq.
(\ref{radap}). One of the basic physical characteristics of the
gravitational bremsstrahlung radiation is that it is independent
of the mass of the radiating particle. Therefore the intensity of
the radiation increases very rapidly with the charge of the body,
and can reach very high values for macroscopic objects. If the
electromagnetic power $P$ from particles moving either on circular
paths or in free fall towards the wormhole is measured, as well as
their velocities $v$ and positions $r$, then the conformal factor
$\psi $ can be obtained from the observational data as
\begin{equation}
\psi \approx k+\sqrt{\frac{3}{2}}\frac{r}{e\gamma ^{2}}\sqrt{P}.
\end{equation}

\section{Discussions and final remarks}\label{secVI}

In the present paper we have considered the dynamics of test
particles in stable circular orbits around static and spherically
symmetric wormholes in conformally symmetric spacetimes.  The
analysis of this problem is important, because, through the
observation of the shifts in the frequency of light  emitted by
high velocity, ultra-relativistic particles rotating around
compact astrophysical objects, it may open the possibility of the
observational detection of these hypothetical general relativistic
objects.

In the case of the conformally symmetric wormholes, the main
result of our analysis is the direct relation between a purely
geometric quantity, i.e., the conformal factor $\psi $, and an
observable quantity, the tangential velocity $v_{\mathrm{tg}}$ of
the test particles moving in stable circular orbits around the
central wormhole. Consequently, all the components of the metric
tensor describing the geometrical properties of the wormhole,
namely, the redshift function $\Phi(r)$ and the form function
$b(r)$, can be expressed in terms of an observable physical
quantity. Moreover, other directly observable quantities, namely,
the shifts in the frequency of light emitted by a particle
rotating around a wormhole can also be expressed in terms of the
conformal factor.

Specific physical properties and characteristics of gravitational
bremsstrahlung emitted by charged particles in geodesic motion in
conformally symmetric wormhole geometries were also explored, and
expressions relating the radiated power and the conformal factor
were found. Thus, it is interesting to note that the conformal
factor $\psi$ can be obtained from the observational data, if the
electromagnetic power from charged particles moving either on
circular paths or in free fall towards the wormhole is measured,
as well as their velocities and positions.

In light of the points emphasized above, the observation of these
shifts and the radiated power from charged particles, through
gravitational bremsstrahlung, could provide a direct observational
test of the wormhole geometry, and implicitly, of the conformal
structure of the space-time. Note that the analysis considered
throughout this work could be generalized to other general
relativistic compact objects and also by imposing a non-static
conformal function $\psi (r,t)$, where a wider variety of exact
solutions may be found. However, this shall be analyzed in a
future work.

\acknowledgments The work of TH was supported by the RGC grant
No.~7027/06P of the government of the Hong Kong SAR. FSNL was
funded by Funda\c{c}\~{a}o para a Ci\^{e}ncia e a Tecnologia
(FCT)--Portugal through the grant SFRH/BPD/26269/2006.

\end{document}